\documentclass[aps,prd,twocolumn,showpacs,floatfix,superscriptaddress,
preprintnumbers,nofootinbib]{revtex4}
\usepackage{graphicx,amssymb,url}

\newcommand{\be}{\begin{equation}}
\newcommand{\ee}{\end{equation}}
\newcommand{\ba}{\begin{eqnarray}}
\newcommand{\ea}{\end{eqnarray}}

\def\lsim{\raise0.3ex\hbox{$\;<$\kern-0.75em\raise-1.1ex\hbox{$\sim\;$}}}
\def\gsim{\raise0.3ex\hbox{$\;>$\kern-0.75em\raise-1.1ex\hbox{$\sim\;$}}}

\def\theta{\vartheta}
\def\ap{\approx}

\begin{document}

\title{B/C ratio and the PAMELA positron excess}

\author{M.~Kachelrie\ss}
\affiliation{Institutt for fysikk, NTNU, Trondheim, Norway}

\author{S.~Ostapchenko}
\affiliation{Institutt for fysikk, NTNU, Trondheim, Norway}
\affiliation{D.~V.~Skobeltsyn Institute of Nuclear Physics,
 Moscow State University, Russia}

\date{November 6, 2012}

\begin{abstract}
We calculate the diffuse intensity of cosmic ray (CR) nuclei and their 
secondaries in the Boron-Carbon group produced by supernova remnants (SNR).
The trajectories of charged particles in the SNR are modeled as a random 
walk in the test particle approximation. Secondary production by CRs 
colliding with gas in the SNR is 
included as a Monte Carlo process, while we use Galprop to account for the 
propagation and interactions of CRs in the Galaxy. 
In the vicinity of a source, we find an approximately constant B/C ratio as 
a function of energy. As a result, the  B/C ratio at Earth does not 
rise with energy, but  flattens instead in the high energy limit.
This prediction can be soon tested by the AMS-2 collaboration.
\end{abstract}


\maketitle

\section{Introduction}

Measurements of the nuclear composition and of the antimatter fraction 
of cosmic rays (CR) are one of the  main tools
 to understand their origin~\cite{galprop}. 
Because the propagation of charged CRs in the turbulent component of the 
Galactic magnetic field is  diffusive up to 
energies $E/Z\lsim 10^{17}$\,eV, properties as the average injection
spectrum of CRs or the diffusion tensor and its energy dependence
have to be determined indirectly. The quality of these measurements
has been improved in the last years considerably, for instance
by the determinations of the intensity of CR protons and He 
performed by the PAMELA collaboration~\cite{pamela}. In particular, the        
PAMELA collaboration found a slope change of the CR spectrum, which 
is difficult to explain in the simplest models~\cite{VJMP}. 
Similar deviations from a pure power-law even below the knee
region were found using gamma-ray observations of selected molecular
clouds in the Gould belt~\cite{NST,KO}.

Previously, the PAMELA collaboration measured the positron fraction in CRs
and found a rapid rise from 10 to 100\,GeV~\cite{PAMELA}. 
Fermi LAT measurements confirmed this behavior in the energy range 
20--200\,GeV~\cite{FermiLAT:2011ab}.  In contrast, 
the antiproton ratio measured by PAMELA declines
above 10\,GeV~\cite{Adriani:2008zq}, consistent with expectations.
Such a rise of the positron fraction is most naturally
explained by the injection of  high-energy positrons by local sources as 
nearby pulsars~\cite{pulsar}: Since electrons loose fast energy, 
the high-energy part of the $e^-+e^+$ spectrum is dominated 
by local sources. Moreover, electromagnetic pair cascades in pulsars 
result naturally in a large positron fraction together with a ``standard'' 
antiproton flux.

In this work we study an alternative explanation for the rising secondary 
fraction which was put forward originally for positrons in Ref.~\cite{B09}: 
Secondaries created by hadronic 
interactions in the shock vicinity participate in the acceleration process 
and therefore it was suggested that they have a flatter 
energy spectrum than primary CRs.  In Ref.~\cite{B09}, it was estimated 
that the resulting positron fraction can explain the PAMELA excess and 
increases up to 50\%, while subsequently this mechanism  was applied to
antiprotons in Ref.~\cite{pp} and to the B/C and Ti/Fe ratios in 
Refs.~\cite{sarkar}.

The present work examines the acceleration and production of nuclei in SNRs.
Following closely in spirit our previous studies~\cite{1,2}, we
use a Monte Carlo (MC) approach calculating the trajectory of each 
particle individually in a random walk picture. This makes it easy to 
include interactions and the production of secondaries. 
As our main result, we predict a flat B/C ratio in the source
vicinity, without a rise at high energies. At Earth, the additional
contribution from SNRs to the secondary fluxes  only flattens the  
B/C ratio but does not lead to a rise in the high energy limit.
 Consequently, an increase in the antiproton-proton ratio
cannot be explained by astrophysical models and could be used as
a signature of DM. This prediction can be tested soon by the AMS-2
collaboration~\cite{AMS2}, which aims at a 1\% accurate measurement
of the flux of different nuclei up to energies of TeV/nucleon~\cite{pohl}.

\section{Simulation procedure}


As we are following closely the approach presented already in 
Refs.~\cite{1,2}, we recall only briefly our methodology.
We model trajectories by a random walk in three dimensions with step 
size $l_0(E)$ determined by an energy-dependent, isotropic diffusion 
coefficient $D$. We assume as usual in the test particle approach
that the diffusion of a nucleus with charge $Ze$ and momentum $p$ 
proceeds in the Bohm regime,
\be
 D = \frac{cl_0}{3} = \frac{cR_L}{3} = \frac{c^2 p}{3ZeB} \,,
\ee
where $B$ denotes the turbulent magnetic field.
Then the mean free path $l_0$ is given by the Larmor radius $R_L$ which 
in turn is determined by $R_L= p/(ZeB)$.  Thus the step-size $l_0$ and the 
time step $\Delta t=l_0/c$ is controlled by $p/Z$.  As a draw-back of our 
Monte Carlo approach, the step-size in the random walk decreases therefore
as $l_0\propto 1/Z$, requiring larger computing time for nuclei with
higher charges $Ze$. 
Therefore we restrict our analysis to relatively light nuclei, concentrating 
on the B/C ratio.

We do not consider any feedback of CRs on the shock or the magnetic field.
As the strength of the magnetic field
is assumed to change from a phase of amplification to damping,
we use two models for the time evolution of the magnetic field $B$:
In the first case, we use a constant turbulent magnetic field 
with $B=1\;\mu$G during all the evolution of the SNR. As the 
second choice, we assume strong field amplification in the early phase,
with $B = 100\;\mu$G before the transition to the  Sedov-Taylor phase 
at $t_\ast=240$\,yr,  and subsequent damping to 
$B = 1/20\:\mu$G at $t>t_\ast$.

For the position $r_{\rm sh}$ and the velocity $v_{\rm sh}$ of the
SNR shock we use the  $n=0$ case of the analytical solutions 
derived in Ref.~\cite{TM99}. These solutions connect smoothly
the ejecta-dominated phase with free expansion $r_{\rm sh}\propto t$
and the Sedov-Taylor stage $r_{\rm sh}\propto t^{2/5}$. The acceleration of 
CRs is assumed to cease after the transition to the radiative
phase at the time $t_{\max}$. We use an age-limited scenario for the
CR escape where most CRs are accumulated downstream and are released
at $t_{\max}$. The SNR is modeled by the following parameters: We choose the
injected mass as $M_{\rm ej} = 4M_\odot$, the mechanical explosion energy 
as $E_{\rm snr} = 5\times 10^{51}$\,erg, and the density of the ISM as    
$n_{\rm ISM} = 2$\,cm$^{-3}$. The end of the Sedov-Taylor phase follows
then as $t_{\max}= 13.000$\,yr. 
We model the injection rate  proportional to the CR pressure~\cite{in2},
\be 
 \dot N \propto R_s^2 v_{\rm sh}^\alpha \delta(E-E_0)\delta(r-r_{\rm sh}) 
\ee
with $\alpha=3$. We choose the isotopes according to the source abundances 
of Ref.~\cite{galprop} and inject isotopes up to $A=20$. 
Lifetimes for radioactive isotopes
are taken from the database~\cite{bnl}.
We use for the simulation of nuclear reactions the cross sections
and decay tables extracted from GALPROP which in turn are based on
Refs.~\cite{sigmas,webber}.

We describe the propagation of cosmic rays in the Galaxy with the
help of the Galprop code \cite{galprop,galprop2}.
More specifically, we consider plain diffusion, neglecting re-acceleration 
process and fixing all relevant parameters, like the normalization and the 
energy-slope of the diffusion coefficient, from a fit to HEAO 
data~\cite{heao} to both
primary and secondary nuclei in the $1\div 30$ GeV range.

\begin{figure}
\includegraphics[width=\linewidth, height=5cm]{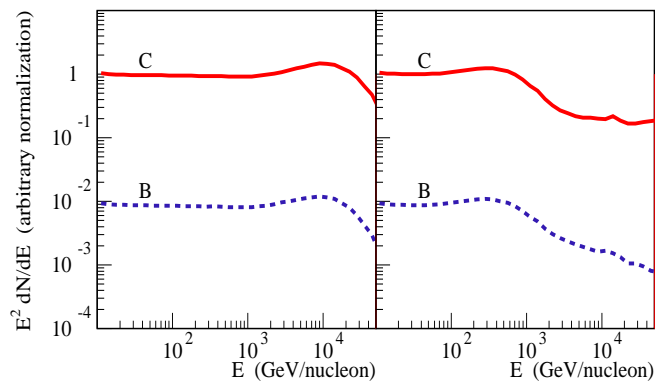}
\caption{
Boron (dashed blue) and Carbon (solid red) energy spectra $E^2dN/dE$ at 
the source as a function of energy per nucleon for two different 
magnetic field models: left panel constant and right panel time-dependent
magnetic field.
\label{fig:CR-source}}
\end{figure}

\begin{figure}
\includegraphics[width=0.8\linewidth]{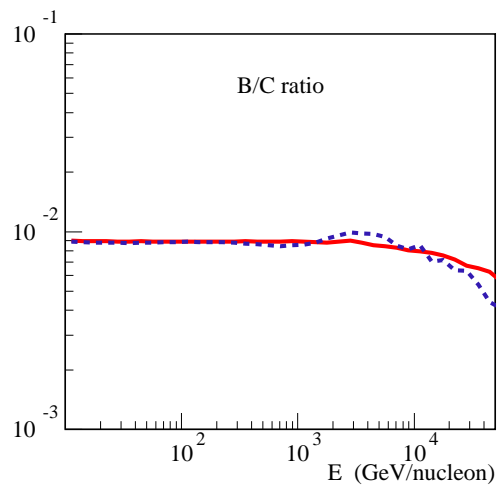}
\caption{
The B/C ratio at the source as a function of energy  per nucleon
for the constant (solid red) and time-dependent (dashed blue) 
magnetic field case.
\label{fig:ratio1}}
\end{figure}

\begin{figure*}
\includegraphics[width=0.8\linewidth]{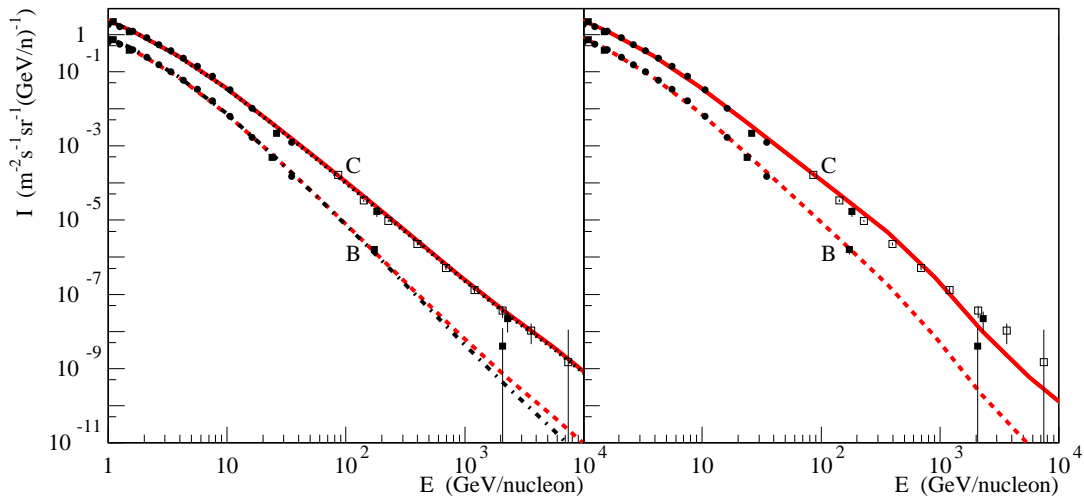}
\caption{
Diffuse intensity of Boron (red dashed) and Carbon (red solid) at the position
of the  Earth for the constant (left panel) and time-dependent (right panel) 
magnetic field case. Additionally, in the left panel the B and C intensity
including only secondary production in the ISM are shown by black dot-dashed 
and black doted lines respectively. Experimental data: 
HEAO \cite{heao} - filled circles, CREAM  \cite{cream} - open squares,  
TRACER  \cite{tracer} - filled  squares.
\label{fig:CR-earth}}
\end{figure*}

\begin{figure}
\includegraphics[width=0.9\linewidth]{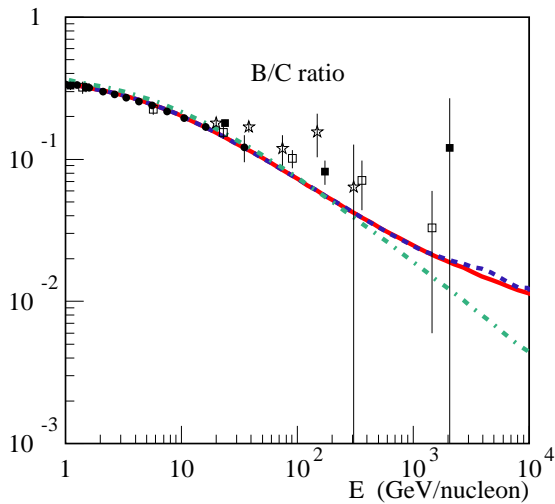}
\caption{
The B/C ratio as a function of energy per nucleon at the position
of the  Earth for the constant (red solid) and time-dependent (blue dashed) 
magnetic field case; the green dot-dashed line shows the  B/C ratio 
including only  secondary production in the ISM .
 Experimental data: HEAO \cite{heao} - filled circles, 
 ATIC~\cite{Panov:2007fe} - open stars,
CREAM  \cite{Ahn:2008my} - open squares, and
 TRACER  \cite{tracer} - filled  squares.
\label{fig:ratio-earth}}
\end{figure}

\section{Numerical results}

In Fig.~\ref{fig:CR-source}, we show the energy dependence of the B
($^{10}{\rm B}+^{11}\!{\rm B}$) and C ($^{12}{\rm C}+^{13}\!{\rm C}$) 
spectra  $E^2dN/dE$ at $t_{\max}$ for the two cases of  magnetic field 
evolution. For a constant magnetic field, 
the spectra $dN/dE$ of both primary and secondary CRs are well 
approximated by an  $E^{-2}$ shape\footnote{The small ``bump'' near the 
acceleration cutoff $E_{\max}\sim {\rm few}\;\times 10^{13}\:Z$eV 
is formed by the small
overdensity of the most energetic CRs in the acceleration zone, as discussed
in more detail in \cite{2}.}. The corresponding B/C ratio shown in 
Fig.~\ref{fig:ratio1} by the solid line remains constant up to the acceleration
cutoff: Because the energy per nucleon $E/A$ in a spallation reaction is
approximately conserved and the simple relation $E/Z\simeq 2E/A$
holds, the rigidity $E/Z$ of the reaction products stays approximately 
unchanged.
Hence, the acceleration process is not influenced by the conversion of, say,
a $^{12}{\rm C}$ into a $^{10}{\rm B}$ nuclei: The secondary nucleus follows 
the same trajectory in a given magnetic field configuration as its parent 
would do, if no conversion took place. As a consequence, the boron and carbon 
spectra as function of energy per nucleon have practically identical shapes,
with their   relative normalization determined approximately by the 
interaction depth in the SNR: The latter is given by
$\tau=ct_{\max}Rn_{\rm ism}\sigma_C\ap 0.01$
assuming $\sigma_{C\to B}\sim 50$\,mbarn as typical value for the fragmentation
cross section of a carbon nuclei into boron.

In the second model for the magnetic field evolution,  the spectral shape 
is more complicated:  Nuclei that were injected early are accelerated up to  
few$\;\times 10^{15}\:Z$eV, while the bulk of CRs injected when the 
turbulent magnetic field is damped has a cutoff  around $10^{12}\:Z$eV.
However, due to the conservation of the rigidity by the spallation process,
the energy spectra of B and C are remarkably similar: The B/C ratio is again
energy-independent and coincides with the one obtained for the case of a 
constant magnetic field---as it should, since the interaction depth is the 
same in the two cases. Thus, our conclusion on the energy-independence of 
the primary
to secondary CR ratios is of general character, being independent of a
particular scenario for the source magnetic field.

As the next step, we calculate the CR propagation in the Galaxy
using the Galprop code, and compare in Fig.~\ref{fig:CR-earth} the 
resulting B and C intensity at Earth with experimental data. For the 
spatial diffusion coefficient in the Galaxy, we used 
$D_{xx}(E)=D_0 (E/10\,{\rm GeV})^{0.65}$,
with $D_0=4.8\times 10^{28}\:{\rm cm}^2/{\rm s}$.
In the left panel of  Fig.~\ref{fig:CR-earth},  we show additionally
the B and C intensity including only secondary production in the interstellar 
medium (ISM), i.e.\ neglecting the SNR contribution to the B intensity.
The SNR contribution to the Boron intensity remains small in the whole 
energy range considered, independent of the models used for the magnetic 
field evolution in the SNR.  At Earth, the
 obtained B/C ratio shown in  Fig.~\ref{fig:ratio-earth} coincide with the
pure propagation results up to $\sim {\rm few} \times 100$\,GeV/nucleon; 
at higher energies the ratios slowly flatten to the per cent level, i.e.\ 
to the respective source value.

\section{Conclusions}%
We calculated the energy spectra of nuclei in the B-C group and their 
secondaries produced in a supernova remnant (SNR) using a simple random 
walk picture. In contrast to a previous prediction \cite{sarkar}
that the B/C ratio rises sharply above energies $\sim $\,TeV/nucleon, 
we found that the ratio  remains approximately constant in the vicinity 
of a source. Our results suggest that the contribution of SNRs to the
intensity of secondary  CRs is subdominant up to very high energies.

In our simulations we used an age-limited escape of CRs. Implementing a 
free-escape boundary may change our results, except if the boundary is 
placed at the distance $\propto E$. In particular, assuming the same 
position of the free-escape boundary for all energies will lead to a softening
of the secondary spectra \cite{Kawanaka:2012xg}, lowering further
the B/C ratio.

This prediction can be soon tested by the AMS-2 collaboration~\cite{AMS2}.
The absence of a rise in the B/C  ratio would support the arguments
of Refs.~\cite{1,2} that reacceleration close to shock fronts in SNRs
does not lead to a change in the
secondary to primary ratios. 

\acknowledgments
We are grateful to Ricard Tom\`as for collaboration in the early
stage of this project, and to Igor Moskalenko for advice on Galprop
and comments on the text.

S.O. acknowledges the support of Norsk Forskningsradet within the program 
Romforskning.




\begin{thebibliography}{00}


\bibitem{galprop}
A.~W.~Strong and I.~V.~Moskalenko,
  Adv.\ Space Res.\  {\bf 27}, 717 (2001);
see also \url{http://galprop.stanford.edu/web_galprop/galprop_home.html}.

\bibitem{pamela}
O.\ Adriani {\it et al.}, Science {\bf 332}, 69 (2011).

\bibitem{VJMP}
A.~E.~Vladimirov, G.~J{\'o}hannesson, I.~V.~Moskalenko and T.~A.~Porter, 
 Astrophys.\ J.\  {\bf 752}, 68 (2012)
[arXiv:1108.1023].


\bibitem{NST}
A.\ Neronov, D.\ V.\ Semikoz, and A.\ M.\ Taylor,
Phys.\ Rev.\ Lett.\ {\bf 108},  051105 (2012).

\bibitem{KO}
M.~Kachelrie{\ss} and S.~Ostapchenko,
  Phys.\ Rev.\ D {\bf 86}, 043004 (2012)
  [arXiv:1206.4705 [astro-ph.HE]].

\bibitem{PAMELA}
  O.~Adriani {\it et al.}  [PAMELA Collaboration],
  Nature {\bf 458}, 607 (2009).


\bibitem{FermiLAT:2011ab} 
  M.~Ackermann {\it et al.}  [Fermi LAT Collaboration],
  Phys.\ Rev.\ Lett.\  {\bf 108}, 011103 (2012)
  [arXiv:1109.0521 [astro-ph.HE]].


\bibitem{Adriani:2008zq}
  O.~Adriani {\it et al.},
  Phys.\ Rev.\ Lett.\  {\bf 102}, 051101 (2009).

\bibitem{pulsar}
 A.~K.~Harding and R.~Ramaty, 
Proc. 20th ICRC, Moscow, {\bf 2}, 92 (1987);
A.~Boulares, 
Astrophys.\ J.\ {\bf 342}, 807 (1989);
F.~A.~Aharonian, A.~M.~Atoyan and H.~J.~V\"olk,
  Astron.\ Astrophys.\  {\bf 294}, L41  (1995).


\bibitem{B09}
P.~Blasi,
  Phys.\ Rev.\ Lett.\  {\bf 103}, 051104 (2009).

\bibitem{pp}
P.~Blasi and P.~D.~Serpico,
  Phys.\ Rev.\ Lett.\  {\bf 103}, 081103 (2009).

\bibitem{sarkar}
P.~Mertsch and S.~Sarkar,
  Phys.\ Rev.\ Lett.\  {\bf 103}, 081104 (2009)
  [arXiv:0905.3152 [astro-ph.HE]];
M.~Ahlers, P.~Mertsch and S.~Sarkar,
  Phys.\ Rev.\  D {\bf 80}, 123017 (2009);
see also 
N.~Tomassetti and F.~Donato,
  Astron.\ Astrophys.\  {\bf 544}, A16 (2012)
  [arXiv:1203.6094 [astro-ph.HE]].


\bibitem{1}
  M.~Kachelrie\ss, S.~Ostapchenko and R.~Tom\`as,
  J.\ Phys.\ Conf.\ Ser.\  {\bf 259}, 012092 (2010)
  [arXiv:1004.1118 [astro-ph.HE]].


\bibitem{2}
  M.~Kachelrie\ss, S.~Ostapchenko and R.~Tom\`as,
  Astrophys.\ J.\  {\bf 733}, 119 (2011)
  [arXiv:1103.5765 [astro-ph.HE]].



\bibitem{AMS2}
V.~Bindi for the AMS-02 collaboration,
``Status of the AMS-02 detector after one year of operation on the International Space Station'' at 
ICHEP2012 - 36th International Conference for High Energy Physics, Melbourne 2012.

\bibitem{pohl}
M.~Pohl at the workshop ``Searching for the sources of Galactic cosmic rays,''
Paris 2012 
\url{http://www.apc.univ-paris7.fr/~semikoz/CosmicRays/CosmicRays/Dec12/pohl.pdf}


\bibitem{TM99}
J.~K.~ Truelove and Ch.~F.~McKee, 
Astrophys.\ J.\ Suppl.\ {\bf 120}, 299 (1994).




\bibitem{in2}
  V.~S.~Ptuskin and V.~N.~Zirakashvili,
  Astron.\ Astrophys.\  {\bf 429}, 755 (2005).


\bibitem{bnl}
\url{http://www.nndc.bnl.gov/nudat2}

\bibitem{sigmas}
S.\ G.\ Mashnik, K.\ K.\ Gudima,   I.~V.~Moskalenko, R.\ E.\ Prael and A.\ J.
Sierk,
   Adv.\ Space Res.\  {\bf 34}, 1288 (2004).

\bibitem{webber}
W.\ R.\ Webber, A.\ Soutoul, J.\ C.\ Kish and J.\ M.\ Rockstroh,
Astrophys.\ J.\ Suppl.\ {\bf 144}, 153 (2003).


\bibitem{galprop2}
I.~V.~Moskalenko and A.~W.~Strong,
 Astrophys.\ J.\  {\bf 493}, 694  (1998); 
A.~W.~Strong and I.~V.~Moskalenko, {\em ibid.} {\bf 509} (1998) 212;
A.~E.~Vladimirov et {\it al.}, 
Comput.\ Phys.\ Commun.\  {\bf 182}, 1156  (2011).


\bibitem{heao}
J.\ J.\ Engelmann, P.\ Ferrando, A.\ Soutoul, P.\ Goret and E.\ Juliusson,
  Astron.\ Astrophys.\  {\bf 233}, 96 (1990).


\bibitem{cream}
H.\ S.\ Ahn et {\it al.},
 Astrophys.\ J.\  {\bf 707}, 593 (2009).


\bibitem{tracer}
A.~Obermeier et {\it al.}, 
Astrophys.\ J.\  {\bf 742},  14 (2011).

\bibitem{Panov:2007fe} 
  A.~D.~Panov et {\it al.}, 
Proc. ``30th International Cosmic Ray Conference,
{\bf 2}, 3 (2008)
  arXiv:0707.4415 [astro-ph].

\bibitem{Ahn:2008my} 
  H.~S.~Ahn et {\it al.}, 
  Astropart.\ Phys.\  {\bf 30}, 133 (2008)
  [arXiv:0808.1718 [astro-ph]].




\bibitem{Kawanaka:2012xg}
  N.~Kawanaka,
  arXiv:1207.0010 [astro-ph.HE].








\end{thebibliography}
\end{document}